\documentclass[aps,a4paper,prl,twocolumn,superscriptaddress,groupedaddress,11pt]{revtex4-2}  
\usepackage{graphicx}  
\usepackage{dcolumn}   
\usepackage{bm}        
\usepackage{amssymb}   
 \usepackage{float}
\usepackage{amsfonts}
\usepackage{gensymb}
\usepackage{amsmath}
\usepackage{upgreek}
\usepackage[cal=RSFS]{mathalfa}
 \usepackage{hyperref}

\hypersetup{
   colorlinks=true,
    linkcolor=blue,
    filecolor=magenta,      
    urlcolor=blue,
}
 
\usepackage{blindtext,tikz}
\usetikzlibrary{calc}
\begin{document}
\title{Existence of Chern Gaps in Kagome Magnets RMn$_6$Ge$_6$ (R = Nd, Sm, Tb, Dy, Ho, Er, Yb, Lu)}

\author{Christopher Sims}
\affiliation{\textit{School of Electrical and Computer Engineering, Purdue University, West Lafayette, IN 47907, USA}}
\date{\today}
\begin{abstract}

{Kagome Chern Magnets are lattices which host a Chern state when they are non-magnetic and a Chern Gap when they become magnetic. These Chern gaps can host fractional Chern Fermions. There has been extensive research in the RMn$_6$Sn$_6$ system which hosts a small Chern gap with many trivial bands surrounding them. RMn$_6$Ge$_6$ is studied theoretically in order to find a more insulating Kagome magnetic with fewer bands near the Chern gap .}
\end{abstract}
\date{\today}
\maketitle

\section{Introduction}

\begin{table*}[t]
\begin{tabular}{|c|c|c|c|c|c|c|c|c|}\hline
No. 			  & Crystal        				& Abbreviation 	  & A     			  & B       & C     & Magnetism & Hubbard U & Hubbard J \\ \hline
{\color{red} N/A} &{\color{red} LaMn$_6$Ge$_6$} & {\color{red} N/A}  & {\color{red} N/A}   & {\color{red} N/A}     & {\color{red} N/A}   & {\color{red} N/A}  & {\color{red} N/A} & {\color{red} N/A}      \\
{\color{red} N/A} & {\color{red} CeMn$_6$Ge$_6$} &{\color{red} N/A}  & {\color{red} N/A}   & {\color{red} N/A}     & {\color{red} N/A}   & {\color{red} N/A}  & {\color{red} N/A} & {\color{red} N/A}      \\
{\color{red} N/A} & {\color{red}  PrMn$_6$Ge$_6$} & {\color{red} N/A}  & {\color{red} N/A}   & {\color{red} N/A}     & {\color{red} N/A}   & {\color{red} N/A}   & {\color{red} N/A} & {\color{red} N/A}     \\
1    				   & NdMn$_6$Ge$_6$ & NMG  & 5.227 & -2.6135 & 8.178 & Ferrimagnetic & 3.1 & 1    \\
{\color{red} N/A} & {\color{red}  PmMn$_6$Ge$_6$} & {\color{red} N/A}  & {\color{red} N/A}   & {\color{red} N/A}     & {\color{red} N/A}   & {\color{red} N/A}   & {\color{red} N/A} & {\color{red} N/A}     \\
2   				  & SmMn$_6$Ge$_6$ & SMG  & 5.208 & -2.604  & 8.155 & Ferrimagnetic   & 3.3 & 1  \\
{\color{red} N/A} & {\color{red}  EuMn$_6$Ge$_6$} & {\color{red} N/A}  & {\color{red} N/A}   & {\color{red} N/A}     & {\color{red} N/A}   & {\color{red} N/A}  & {\color{red} N/A} & {\color{red} N/A}      \\
{\color{red} N/A} & {\color{red}  GdMn$_6$Ge$_6$} & {\color{red} N/A}  & {\color{red} N/A}   & {\color{red} N/A}     & {\color{red} N/A}   & {\color{red} N/A}  & {\color{red} N/A} & {\color{red} N/A}      \\
3   				  & TbMn$_6$Ge$_6$ & TMG  & 5.079 & -2.5395 & 8.16  & Ferrimagnetic  & 5 & 1  \\
4   				  & DyMn$_6$Ge$_6$ & DMG  & 5.178 & -2.589  & 8.143 & Ferrimagnetic  & 5 & 1  \\
5   				  & HoMn$_6$Ge$_6$ & HMG  & 5.178 & -2.589  & 8.17  & Ferrimagnetic    & 4.9 & 1 \\
6   				  & ErMn$_6$Ge$_6$ & EMG  & 5.172 & -2.586  & 8.139 & Ferrimagnetic   & 4.2 & 1   \\
{\color{red} N/A} & {\color{red}  TmMn$_6$Ge$_6$} & {\color{red} N/A}  & {\color{red} N/A}   & {\color{red} N/A}     & {\color{red} N/A}   & {\color{red} N/A}     & {\color{red} N/A} & {\color{red} N/A}   \\
7   				  & YbMn$_6$Ge$_6$ & YMG  & 5.186 & -2.593  & 8.132 & Ferrimagnetic   & 3 & 0.67  \\
8   				   & LuMn$_6$Ge$_6$ & LMG  & 5.16  & -2.58   & 8.122 & Ferrimagnetic & 5.5 & 0.95 \\\hline
\end{tabular}
  	\caption{\textbf{Lattice Parameters:} Lattice parameters and magnetism type for all rare earth RMn$_6$Ge$_6$ materials}
\label{params}
\end{table*}

\indent Kagome magnets\cite{Guo2009,Nayak2016,Yang2017,Kang2019,Ghimire2020} have gain a recent interest in the field of condensed matter physics due to the fact that they may host fractional Chern Fermions in their band gaps\cite{Haldane1988,Thonhauser2006,Jotzu2014,Skirlo2015}. The first material that was discovered to hose this property in the 166-Kagome system was TbMn$_6$Sn$_6$\cite{Yin2020}, this was due to the fact that the orbital magnetism was out of plane, making it easy to study cleaved planes in the layered compound. In addition TbMn$_6$Sn$_6$ and other Kagome magnets exhibit the anamolous hall effect \cite{Chen2021,Dhakal}, which is a result of the Chern invariant existing in this system.

A large interest in the 166-Kagome magnet system\cite{Idrissi1991,Venturini1993,Zhang2001,Shaoying2001,Mazet2006,Mazet2007,Ghimire2020a,Ma2021} is garnered due to the fact that there can exist a Chern gap in the magnetic phase whose study\cite{Halperin2011,Bartolomei2020} can elucidate the nature of non-abelian particles in confined systems\cite{Sarma2005,Stern2013}. There has been extensive research in the RMn$_6$Sn$_6$ system which hosts a small Chern gap with many trivial bands near the fractional Chern gapped states, however, there has been little interest in other compounds in the 166-Kagome system\cite{Sims2021}.

Fractional Chern Fermions ($\mathbb{C}_{\mathcal{F}}$) are non-abelion particles which on the surface of the 166-Kagome magnet system. Non-abelion anyonic particles are nontrivial under braiding operations, picking up a phase\cite{Nayak2008}. These particles exist in fractional hall states in confined 2D fractional hall insulators\cite{Nakamura2020}. Fractional Chern Fermions that are confined to the surface of a Kagome magnet are also non-abelion particles. While fractional aynonic statistics have not been observed in a Chern magnet, the associated Chern gapped states have been observed\cite{Yin2020}.

This work presents a theoretical study of the Chern gaps of Kagome lattices' RMn$_6$Ge$_6$ (R = Nd, Sm, Tb, Dy, Ho, Er, Yb, Lu) whose Fermi surface electronic structures are unique. Typically, materials with similar lattice constants and chemical composition exhibit similar band structures. However, RMn$_6$Ge$_6$ type Kagome lattices, there are a few large differences, this is can be due to the dependence of correlation effects in heavy Fermion rare earth elements.

\begin{figure*}[ht]
 \centering
 \includegraphics[width=1.0\textwidth]{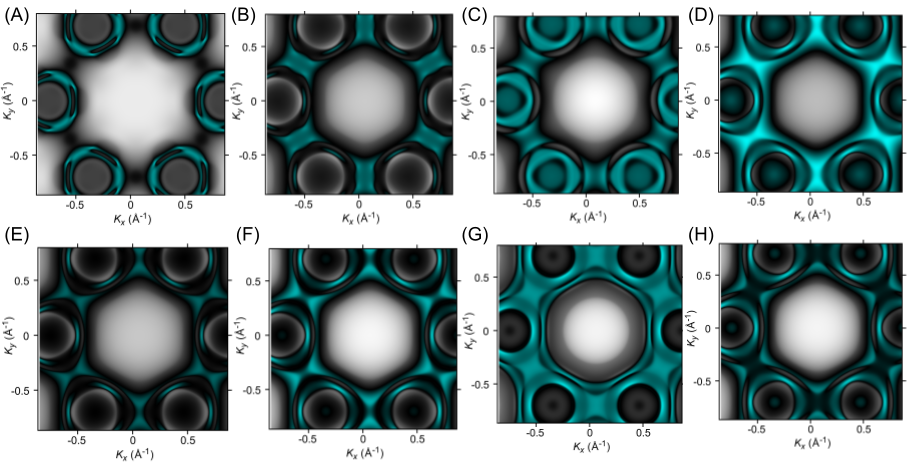}
  	\caption{\textbf{Fermi Surface of RMn$_6$Ge$_6$:} (A) NdMn$_6$Ge$_6$ (NMG) (B) SmMn$_6$Ge$_6$ (SMG) (C) TbMn$_6$Ge$_6$ (TMG) (D) DyMn$_6$Ge$_6$ (DMG) (E) HoMn$_6$Ge$_6$ (HMG) (F) ErMn$_6$Ge$_6$ (EMG) (G) YbMn$_6$Ge$_6$ (YMG) (H) LuMn$_6$Ge$_6$ (LMG)}
\label{ARC}
  \end{figure*}

\section{Methods}
RMn$_6$Ge$_6$ (R = Nd, Sm, Tb, Dy, Ho, Er, Yb, Lu) crystallizes in the Kagome HfFe$_6$Ge$_6$-type structure (space group No. 191, P6/mmm). Rare earth compounds LaMn$_6$Ge$_6$, CeMn$_6$Ge$_6$, PrMn$_6$Ge$_6$, PmMn$_6$Ge$_6$, EuMn$_6$Ge$_6$, GdMn$_6$Ge$_6$, and TmMn$_6$Ge$_6$ do not minimize in the calculcations and have not been found to crystalize utilizing normal crystal growth techniques. As opposed to calculating the minimized electronic structure, experimentally obtained lattice parameters are obtained. The following abbreviations are used throughout this work: NdMn$_6$Ge$_6$ (NMG), SmMn$_6$Ge$_6$ (SMG), TbMn$_6$Ge$_6$ (TMG), DyMn$_6$Ge$_6$ (DMG), HoMn$_6$Ge$_6$ (HMG), ErMn$_6$Ge$_6$ (EMG), YbMn$_6$Ge$_6$ (YMG), LuMn$_6$Ge$_6$ (LMG).

The band structure calculations were carried out using the density functional theory (DFT) program Quantum Espresso (QE)\cite{wannier90}, with the generalized gradient approximation (GGA)\cite{pbegga} as the exchange correlation functional. Projector augmented wave (PAW) pseudo-potentials were generated utilizing PSlibrary. Pre-processing for DFT was conducted with the CIF2WAN interface \cite{Sims2020a}. The energy cutoff was set to 100 Ry and the charge density cutoff was set to 700 Ry for the plane wave basis with a k-mesh of 25$\times$25$\times$25.  All of ther parameters used for the calculations are listed in Table \ref{params}. After convergence of the self consistent (SCF) calculation was completed, the Wannier tight binding Hamiltonian was generated from the non-self consistent calculation (NSCF) with Wannier90\cite{wannier90} via the P2W tool in QE. The surface spectrum \cite{sancho1985} was calculated with Wannier Tools \cite{wtools}. For all calculations in this paper, only SOC + ferrimagnetism (with a driving magnetic field) is considered. Chern ($\mathbb{C}$) number analysis was conducted with the Wilson loop algorithm in Wannier Tools.

\begin{figure*}[ht]
 \centering
 \includegraphics[width=1\textwidth]{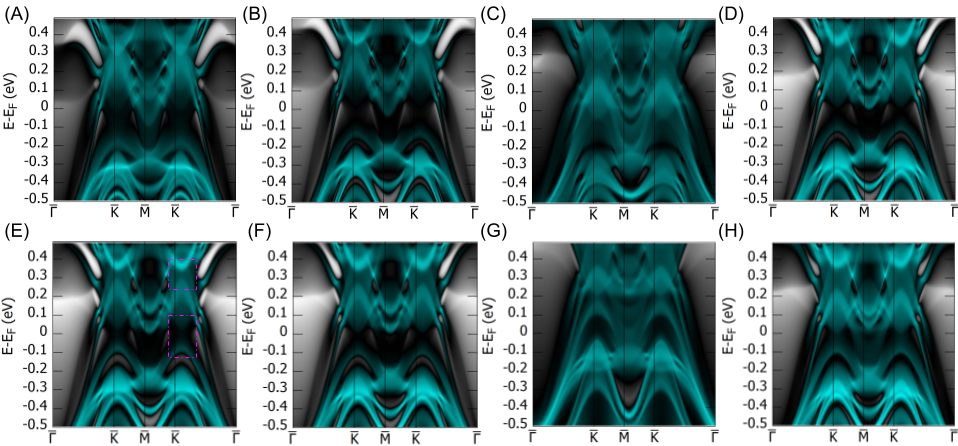}
  	\caption{\textbf{Surface states along the $\mathrm{\overline{\Gamma}}$-$\mathrm{\overline{K}}$-$\mathrm{\overline{M}}$-$\mathrm{\overline{K}}$-$\mathrm{\overline{\Gamma}}$ line:} (A) NdMn$_6$Ge$_6$ (NMG) (B) SmMn$_6$Ge$_6$ (SMG) (C) TbMn$_6$Ge$_6$ (TMG) (D) DyMn$_6$Ge$_6$ (DMG) (E) HoMn$_6$Ge$_6$ (HMG) (F) ErMn$_6$Ge$_6$ (EMG) (G) YbMn$_6$Ge$_6$ (YMG) (H) LuMn$_6$Ge$_6$ (LMG)}
\label{SS}
 \end{figure*}

\section{Results and Discussion}
RMn$_6$Sn$_{6}$ is composed of stacked layers of  RGe$_2$ and GeMn$_3$, with an intermediate Ge$_2$ layer. By viewing the (001) surface of a 2x2 lattice, the Kagome structure can be seen where there is a hexagonal ring composed of Mn-Sn atoms around the central Ho atom at the conventional cell's center and triangles with a central Sn atom surrounded by 3 Mn atoms at the cell's corners. When the system becomes magnetic, R and Mn magnetize to the C-axis in a ferromagnetic configuration. When magnetism is added, the gap widens significantly at the K point to be about 0.3 eV. In addition, the Dirac cone becomes gaped and forms a Chern gaped Dirac cone, which is confirmed by Wilson loop analysis.

Firstly, the Fermi surface arc states near the Chern gap are examined [Figure \ref{ARC}], for all of the surface arc states the Fermi level is chosen for all calculations which exists near the Chern gapped states. The hexagonal structure of the Kagome crystal can be seen with triangular like features around the K points. Interestingly the light rare-earth elements of Nd and Sm [Figure \ref{ARC}(A,B)] Show the most distinct Chern gap states with few trivial bands around the band gap compared to the other materials. This is attributed to there being less valence electrons contributing to the bands near the Chern gap. The other systems show a trivial state [Figure \ref{ARC}(C-H)] which obfuscates the Chern gap states, this is a result of larger Hubbard interactions and more electrons in the valence bands. Wilson loop analysis shows that only a few bands in the the respective RMG systems have a Chern invariant (see supplementary).

In order to further elucidate the nature of the Chern gap states the surface states are exampled in the $\mathrm{\overline{\Gamma}}$-$\mathrm{\overline{K}}$-$\mathrm{\overline{M}}$-$\mathrm{\overline{K}}$-$\mathrm{\overline{\Gamma}}$ line [Figure \ref{SS}]. The surface state calculation show that around the Fermi level there are valence bands that hybridize with the conduction bands to form one of the Chern gap states. The magenta boxes in Figure \ref{SS}(E) outline the multi-stack nature of the Chern gaps in the HMG system The multi-stack Chern states are less prevalent in TMG and YMG [Figure \ref{SS}(C,G)]. It is unknown the cause of why this is the case and warrants further investigation. For the other crystals in the RMG system it is seen that the energy dispersion of the hybridized Chern gap near the Fermi level is about 2.5 eV. This shows that the rare-earth element does not contribute to the size of the Chern gap. Upon viewing a larger energy window, multiple secondary Chern gaps can be seen as far as 1 eV from the Fermi level (see supplementary).

\section{Conclusion}
In conclusion, there exists clear Chern gaps in the RMn$_6$Ge$_{6}$ Kagome magnet system. The nature of how the bands evolve with rare-earth elements are examined. This provides further insight into how Chern gaps evolve in the Kagome Chern magnet system. This work provides interest into the futher study of the RMG system in order study Chern states. In addition, this work promotes the use of band engineering in the highly tunable Kagome magnet system in order to create an ideal Chern gap state.

 \section{Acknowledgments}
 \vspace{-4mm}
C.S. Acknowledges support from the GEM Fellowship and the Purdue Engineering ASIRE Fellowship.
 
 Correspondence and requests for materials should be addressed to C.S.
 (Email: Sims58@Purdue.edu)

 \section{Supplementary}
  \vspace{-4mm}
See supplementary for further details on DFT calculations, topological invariant calculations and bulk band structures of RMn$_6$Ge$_{6}$.

%

\end{document}